\documentclass[pra,preprint,showpacs]{revtex4}
\usepackage{graphicx}
\usepackage{amssymb}
\usepackage{subfigure}
\usepackage{color}

\begin{document}

\title{Ultrashort spatiotemporal optical solitons in quadratic nonlinear media: generation of line and lump solitons from few-cycle input pulses}
\author{Herv{\'e} Leblond$^{1}$, David Kremer$^{1}$, and  Dumitru Mihalache$^{2,3}$ }
\affiliation{
$^{1}$Laboratoire POMA, CNRS FRE 2988,
 Universit\'e d'Angers, 2 Bd. Lavoisier, 49045 Angers Cedex 01, France\\
$^{2}$Horia Hulubei National Institute for Physics and Nuclear Engineering (IFIN-HH),
407 Atomistilor, Magurele-Bucharest, 077125, Romania\\
$^{3}$Academy of Romanian Scientists, 54 Splaiul Independentei, Bucharest 050094, Romania}

\begin{abstract}

By using a powerful reductive perturbation technique, or a multiscale analysis, a 
generic Kadomtsev-Petviashvili evolution equation governing the propagation of femtosecond spatiotemporal optical solitons in quadratic nonlinear media beyond the slowly-varying envelope approximation is put forward. 
Direct numerical simulations show the formation, from adequately chosen few-cycle input pulses, of both stable line solitons (in the case of a quadratic medium with normal dispersion) and of stable lumps (for a quadratic medium with anomalous dispersion). 
Besides, a typical example of the decay of the perturbed unstable line soliton into stable lumps for a quadratic nonlinear medium with anomalous dispersion is also given.

\end{abstract}

\pacs{42.65.Tg, 42.65.Re, 05.45.Yv}

\maketitle

\section{Introduction}

Over the past two decades, the development of colliding-pulse mode-locked lasers and self-started solid-state lasers has been crowned with the generation of few-cycle pulses (FCPs) for wavelengths from the visible to the near infrared (for a comprehensive review see Ref. \cite{BK}).
Presently, there is still a great deal of interest in the study of propagation characteristics of these FCPs in both linear and nonlinear media. 
Thus the comprehensive study of intense optical pulses has opened the door to a series of applications in various fields such as light matter interaction, high-order harmonic generation, extreme nonlinear optics \cite{Wegener}, and attosecond physics \cite{atto1,atto2}. 
The theoretical studies of the physics of FCPs concentrated on three classes of main dynamical models: (i) the quantum approach \cite{tan08,ros07a,ros08a,naz06a}, (ii) the refinements within the framework of the slowly varying envelope approximation (SVEA) of the nonlinear Schr\"{o}dinger-type envelope equations \cite{Brabek_PRL,Tognetti,Voronin,Kumar}, and non-SVEA models \cite{quasiad,leb03,igor_jstqe,igor_hl,interaction}.  
In media with cubic optical nonlinearity (Kerr media) the physics of (1+1)-dimensional FCPs  can be adequately described beyond the SVEA by using different dynamical models, such as the modified Korteweg-de Vries (mKdV) \cite{quasiad}, sine-Gordon (sG) \cite{leb03,igor_jstqe}, or mKdV-sG equations \cite{igor_hl,interaction}.
The above mentioned partial differential evolution equations admit breather solutions, which are suitable for describing the physics of few-optical-cycle solitons. 
Moreover, a non-integrable generalized Kadomtsev-Petviashvili (KP) equation \cite{KP} (a two-dimensional version of the mKdV model) was also put forward for describing the (2+1)-dimensional few-optical-cycle spatiotemporal soliton propagation in cubic nonlinear media beyond the SVEA~\cite{igor1,igor_matcom}. 

Notice that of particular interest for the physics of the (1+1)-dimensional FCPs is the mKdV-sG equation; thus by using a system of two-level atoms, it has been shown in Ref. \cite{igor_hl} that the propagation of ultrashort pulses in Kerr optical media
is fairly well described by a generic mKdV-sG equation. It is worthy to mention that this quite general model was also derived and studied in Refs. ~\cite{saz01b,bugay}. However, another model equation for describing the physics of FCPs, known as the short-pulse equation (SPE) has been introduced, too \cite{sch04a,chung05,sakovich05,sakovich06} and some vectorial versions of SPE have been also investigated \cite{pie08a,ami08a,kim08a}.  
Other kind of SPE containing additional dispersion term has been introduced in Ref. \cite{kozlov97}. A 
multi-dimensional version of the SPE was also put forward \cite{bespalov} and  
self-focusing and pulse compression have been studied, too \cite{berkovsky}. 
Notice that the SPE model has been considered again in its
vectorial version and the pulse self-compression and FCP soliton propagation have been investigated in detail \cite{skobelev}. The main dynamical (1+1)-dimensional FCP models mentioned above have been revisited recently  \cite{LM_PRA_2009}; it was thus proved that the generic dynamical model based on the mKdV-sG partial differential equation was able to retrieve the results reported so far in the literature, and so demonstrating
its remarkable mathematical capabilities in describing the physics of (1+1)-dimensional FCP optical solitons \cite{LM_PRA_2009}. 

As concerning the extension of these studies to other relevant physical settings, it was shown recently that a FCP launched in a quadratically nonlinear medium may result in the formation of a (1+1)-dimensional half-cycle soliton (with a single hump) and without any oscilatting tails \cite{kdvopt}. It was proved that the FCP soliton propagation in a quadratic medium can be adequately described by a KdV equation and not by mKdV equation. Notice that in 
Ref. \cite{kdvopt} it was considered a quadratic nonlinearity for a single wave (frequency), and that no effective third order nonlinearity was involved (due to cascaded second order nonlinearities). This is in sharp contrast with the nonlinear propagation of standard quadratic solitons 
(alias two-color solitons) within the SVEA where two different waves (frequencies) are involved, namely a fundamental frequency and a second harmonic \cite{q1}-\cite{q8}.  

As mentioned above, there are only a few works devoted to the study of (2+1)-dimensional few-optical-cycle solitons, where an additional spatial transverse dimension is incorporated into the model; see e.g., Refs. \cite{igor_jstqe,igor1,igor_matcom}, where the propagation of few-cycle optical pulses in a collection of two-level atoms was investigated beyond the traditional SVEA. The numerical simulations have shown that in certain conditions, a femtosecond pulse can evolve into a stable few-optical cycle spatiotemporal soliton \cite{igor_jstqe,igor1,igor_matcom}. 

The aim of this paper is to derive a generic partial differential equation describing the dynamics of (2+1)-dimensional spatiotemporal solitons in quadratic nonlinear media beyond the SVEA model equations.
Our study rely on the previous work \cite{kdvopt} where the theory of half-cycle (1+1)-dimensional optical solitons in quadratic nonlinear media was put forward.
This paper is organized as follows. In Sec. II we derive the generic KP equation governing the propagation of femtosecond spatiotemporal solitons in quadratic nonlinear media beyond the SVEA. To this aim  we use the powerful reductive perturbation technique up to sixth order in a small parameter $\varepsilon$. We have arrived at the conclusion that when the resonance frequency is well above the inverse of the typical pulse width, which is of the order of a few femtoseconds, the long-wave approximation leads to generic KP I and KP II evolution equations. 
These evolution equations and their known analytical solutions, such as line solitons and lumps are briefly discussed in Sec. III.
Direct numerical simulations of the governing equations are given in Sec. IV, where it is shown the generation of both stable line solitons (for the KP II equation) and of stable lumps (for the KP I equation), and a typical example of the decay of the perturbed unstable line soliton of the KP I equation into stable lumps is also provided.
Section V presents our conclusions. 
 
\section{Derivation of the KP equation}

We consider the same classical and quantum models developed in Ref. \cite{kdvopt}, but with an additional transverse spatial dimension (we call it the $y$-axis).
In the quantum mechanical approach,
the medium composed by two-level atoms is treated using the density-matrix formalism and the dynamics of electromagnetic field is governed by the Maxwell-Bloch equations.
Thus the evolution of the electric field component $E$ along the polarization direction of the wave (the $x$-axis) is described by the (2+1)-dimensional Maxwell equations, which reduce to
\begin{equation}
\left(\partial_y^2+\partial_z^2\right)E=\frac1{c^2}\partial^2_t\left(E+4\pi P\right),\label{max}
\end{equation}
where $P$ is the polarization density.

\paragraph{Classical model.}
First we consider a classical model of an elastically bounded electron oscillating along the polarization
direction of the wave. The wave itself propagates in the $z$-direction. 
The evolution of the position $x$ of an atomic electron is described by an
anharmonic oscillator (here damping is neglected)
\begin{equation}
\frac{d^2 x}{dt^2}+\Omega^2x+a x^2=\frac{-e}m E.\label{elast}
\end{equation}
The parameter $a$ measures the strength of the quadratic nonlinearity.
The polarization density is $P=-Nex$, where $N$ is the density of atoms.
Here $\Omega$ is the resonance frequency, $(-e)$ is the electron charge, $m$ is its mass,  and $E$ is the electric field component along the $x$-axis. 

\paragraph{Quantum model.}
We consider a set of two-level atoms with the Hamiltonian
\begin{equation}
H_0=\hbar\left(\begin{array}{cc}\omega_a&0\\0&\omega_b\end{array}\right),
\end{equation}
where $\Omega=\omega_b-\omega_a>0$ is the frequency of the transition.
The evolution of the electric field $E$ % (we restrict to one field component for the sake of simplicity)
 is described by the wave equation (\ref{max}). The light propagation is
 coupled with  the medium by means of a dipolar electric momentum $\mu$ directed along the same direction
 $x$ as the electric field, according to
 \begin{equation}
 H=H_0-\mu E,\label{eq17}
 \end{equation}
 and the polarization density $P$ along the $x$-direction is
 \begin{equation}
 P=N \mathrm{Tr}\left(\rho\mu\right),\label{eq18}
 \end{equation}
 where $N$ is the volume density of atoms, and $\rho$ is the density matrix.
 The density-matrix evolution equation (Schr{\"o}dinger equation) is written as 
 \begin{equation}
 i\hbar\partial_t\rho=\left[H,\rho\right]+{\cal R},\label{schr}\label{eq19}
 \end{equation}
where  ${\cal R}$ is a phenomenological relaxation term. 
In the case of cubic nonlinearity,
it has been shown that it was negligible~\cite{leb03}, and it will be neglected here as well.
  
In the absence of permanent dipolar  momentum,
\begin{equation}
\mu=\left(\begin{array}{cc}0&\mu\\\mu^\ast&0\end{array}\right)
\end{equation}
is off-diagonal, and the quadratic nonlinearity is zero.

The quadratic nonlinearity can be phenomenologically accounted for as follows:
it corresponds to a deformation of the electronic cloud induced by the field $E$,
hence  to a dependency of the energy of the excited level $b$ with respect to
$E$, i.e., to a Stark effect. The nonlinearity is quadratic if this dependency is linear, as follows
\begin{equation}
\omega_b\longrightarrow\omega_b-\alpha E.
\end{equation}
This contribution can be included  phenomenologically in the Maxwell-Bloch equations by replacing
the free Hamiltonian $H_0$ with
\begin{equation}
H_0-\alpha E\left(\begin{array}{cc}0&0\\0&1\end{array}\right).\label{eqh0p}
\end{equation}

\paragraph{Scaling.}

Transparency implies that the characteristic frequency $\omega_w$ of the considered
radiation (in the optical range)  strongly differs from  the resonance frequency $\Omega$ of the atoms, hence it can be much higher or much lower.
We consider here the latter case, i.e., we assume that $\omega_w$ is much smaller than $\Omega$.
This motivates the introduction of the slow variables
\begin{equation}
\tau=\varepsilon\left(t-\frac zV\right),\quad
\zeta=\varepsilon^3z,\quad
\eta=\varepsilon^2 y,\label{scal}
\end{equation}
$\varepsilon$ being a small parameter.
The delayed time $\tau$ involves propagation at some speed $V$ to be determined.
It is assumed to vary slowly in time, according to the assumption  $\omega_w\ll\Omega$.
The pulse shape described by the variable $\tau$ is expected to evolve slowly in time, the
corresponding scale being that of variable $\zeta$.
The transverse spatial variable $y$ has an intermediate scale as usual in KP-type expansions \cite{tutorial}.

A weak amplitude assumption is needed in order that the nonlinear effects arise at the same propagation distance scale
as the dispersion does. Next we use the reductive perturbation method as developed in Ref. \cite{tutorial}. To this aim we expand the electric field $E$ as power series of a small parameter $\varepsilon$:
\begin{equation}
E=\varepsilon^2E_2+\varepsilon^3E_3+\varepsilon^4E_4+\ldots,
\end{equation}
as in the standard KdV-type expansions~\cite{tutorial}.
The polarization density $P$ is expanded in the same way.

\paragraph{Order by order resolution}
The computation is exactly the same as in \cite{kdvopt} up to order  $\varepsilon^4$ in the Schr\"odinger equation (\ref{schr})
or classical dynamical equation (\ref{elast}), and up to order $\varepsilon^5$ in the wave equation (\ref{max}).
The fourth order polarization density $P_4$ has the same expression as in \cite{kdvopt}.
Then 
the wave equation (\ref{max}) at  order  $\varepsilon^6$ yields the evolution equation
\begin{equation}
\partial_\zeta\partial_\tau E_2=A\partial_\tau^4 E_2+B\partial_\tau^2 \left( E_2\right)^2+\frac V2\partial_\eta^2E_2,\label{kdvn1}
\end{equation}
which is a generic KP equation.

The coefficients $A$ and $B$ in the above (2+1)-dimensional evolution equation are
\begin{equation}
A=\frac{4\pi VNe^2}{2mc^2\Omega^4},\quad B=\frac{-4\pi VaNe^3}{2m^2c^2\Omega^6},
\end{equation}
in the classical model, and respectively,
\begin{equation}
A=\frac{4\pi N|\mu|^2}{nc\hbar \Omega^3},\quad
B=\frac{-4\pi N\alpha |\mu|^2}{nc\hbar^2\Omega^2},
\end{equation}
in the corresponding quantum model.

Notice that the dispersion coefficient $A$ and the nonlinear coefficient $B$ can be written in a general form as
\begin{equation}
A=\frac16\left.\frac{d^3k}{d\omega^3}\right|_{\omega=0}
=\frac1{2c}\left.\frac{d^2n}{d\omega^2}\right|_{\omega=0},
\label{disp1}
\end{equation}
\begin{equation}
B=\frac{-2\pi}{nc}\left.\chi^{(2)}(2\omega;\omega,\omega)\right|_{\omega=0},\label{coefb}
\end{equation}
where $n$ is the refractive index of the medium and $\omega$ is the wave pulsation (for a detailed discussion of the 
corresponding expressions of the dispersion and nonlinear coefficients in the case of cubic nonlinear media see, e.g., Ref. \cite{leb03}). The velocity is obviously $V=c/n$.

\section{Two KP equations: KP I and KP II}

The KP equation (\ref{kdvn1}) is reduced to 
\begin{equation}
\partial_Z\partial_T u=\partial_T^4 u+\partial_T^2 u^2+\sigma\partial_Y^2u,\label{kpr}
\end{equation}
where $\sigma=\mbox{sgn}( A)$, 
by the change of variables
\begin{equation}
\zeta=Z,\quad\tau=A^{1/3} T,\quad \eta=|A|^{1/6}\sqrt{\frac V2}Y,\quad E_2=\frac{A^{1/3}}B u.
\end{equation}

For our starting models (both the classical and quantum ones), the dispersion coefficient $A$ is positive. It is the
case of the normal dispersion. Hence $\sigma=+1$ and Eq. 
(\ref{kpr}) is the so-called KP II equation.
Both KP I and KP II are completely integrable by means of the inverse scattering transform (IST)
method~\cite{ablo2}, however the mathematical properties of the solutions differ.
KP II admits stable {\it line solitons}, but not stable localized 
soliton solutions.

The line soliton has the analytical expression 
\begin{equation}
E_2=\frac{6p^2A}B\mbox{sech}^2\left \{p\left[\tau+a \eta+\left(\frac{a^2V}2+4p^2A\right)\zeta\right]\right \}\label{kdvsoli}
\end{equation}
where $a$ and $p$ are arbitrary constants and $A$, $B$ have the expressions given in the previous Section.

\begin{figure}\begin{center}
\includegraphics[width=7cm]{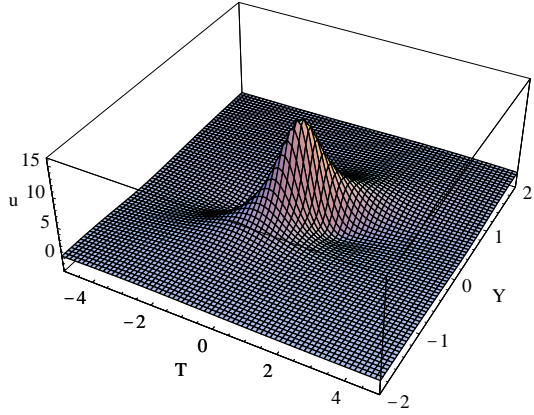}
\caption{(Color online) The lump solution of the KP I equation. Here the parameters are
$A=-1$, $B=-1$, $V=1$, $p_r=-1$, and $p_i=2$.}
\label{lump}
\end{center}
\end{figure}

If Eq. (\ref{kdvn1}) is generalized to other situations using the expressions (\ref{disp1})-(\ref{coefb}) for the dispersion and nonlinear coefficients, and if we assume
an anomalous dispersion, i.e., $A<0$, then $\sigma=-1$ and  Eq. 
(\ref{kpr}) becomes
the so-called KP I equation, for which the above line soliton is not stable, but which admits stable {\it lumps} \cite{sat79,bou97b}
as
\begin{equation}
E_2=\frac{12A}B\frac{\left[-\left(t'+p_r y'\right)^2+p_i^2{y'}^2+3/p_i^2\right]}
{\left[\left(t'+p_r y'\right)^2+p_i^2{y'}^2+3/p_i^2\right]^2},\label{explump}
\end{equation}
with
\begin{equation}
t'=\tau+\left(p_r^2+p_i^2\right)A\zeta,\label{xpr}
\end{equation}
\begin{equation}
y'=\eta\sqrt{\frac{-2A}V}-2p_r A\zeta,\label{ypr}
\end{equation}
where $p_r$ and $p_i$ are arbitrary 
real constants. The analytical lump solution written above is plotted in Fig. \ref{lump}.

The size of the lump can be evaluated as follows.  Assume for the sake of simplicity
that the lump exactly propagates along $z$, 
i.e., that $p_r=0$. Then Eq. (\ref{explump}) has the form
\begin{equation}
%E_2\propto\left(\frac{{t'}^2}{{w_t}^2}+\frac{{y'}^2}{{w_y'}^2}+1\right)^{-2}, 
E_2\propto\left(\frac{{t'}^2}{3/p_i^2}+\frac{{y'}^2}{3/p_i^4}+1\right)^{-2},
\end{equation}
which gives the normalized transverse width and duration as % longitudinal widths as
%\begin{equation}
%w_y'=\frac{\sqrt3}{p_i^2},\quad
%w_t'=\frac{\sqrt3}{p_i},
%\end{equation}
$w_y'=\sqrt3/p_i^2$ and $w_t'=\sqrt3/{p_i}$, respectively,
while the scaled peak amplitude of the lump is
$E_{2max}={4Ap_i^2}/B$.
Using Eqs. (\ref{xpr})-(\ref{ypr}) and the scaling (\ref{scal}), we get the corresponding width and duration
in physical units, as
\begin{equation}
w_y=\frac{\sqrt3}{p_i^2\varepsilon^2}\sqrt {\frac V{-2A}},\quad
w_t=\frac{\sqrt3}{p_i\varepsilon},
\end{equation}
while the peak amplitude is $E_m=\varepsilon^2E_{2max}$ .
Using expressions (\ref{disp1}) and (\ref{coefb}) of the coefficients $A$ and $B$, % and taking the longitudinal width or duration $w_t$ as a parameter,
 we obtain
\begin{equation}
w_y=\frac{w_t^2c}{\sqrt{-3nn''}},\quad
E_m=\frac{-3nn''}{\pi\chi^{(2)}w_t^2},
\end{equation}
where $n''$ and $\chi^{(2)}$ are shortcuts for $\left.{d^2n}/{d\omega^2}\right|_{\omega=0}$ and $\left.\chi^{(2)}(2\omega;\omega,\omega)\right|_{\omega=0}$, respectively.
Hence the  transverse width of the lump decreases as the pulse duration does and the 
peak amplitude increases.
However, the initial assumptions and scaling (\ref{scal})  imply that the field varies much slower in the $y$-direction than longitudinally, hence the width of the lump should, in principle, be large with respect to its length. If this assumption is not satisfied, the
validity of the KP model is questionable.

\section{Numerical simulations of KP I and KP II equations}

The KP equation (\ref{kpr}) is solved by means of the fourth order Runge Kutta exponential time differencing (RK4ETD)
scheme \cite{cox02}. It involves one integration with respect to $\tau$. The inverse derivative is computed by means of a 
Fourier transform, which implies that the integration constant is fixed so that the mean value of the inverse derivative
is zero, but also that the linear term is replaced with zero, i.e.,  the mean value of the function $\partial_y^2u$ is set to zero.
For low frequencies, the coefficients of the RK4ETD scheme are computed by means of series expansions, to avoid catastrophic consequences of limited numerical accuracy.

\subsection{KP II}

For a normal dispersion ($n''>0$), as in the case of the two-level model above, Eq. (\ref{kdvn1}) is KP II.
Then,
starting from an %adequately chosen
 input in the form of a Gaussian plane wave packet
\begin{equation}
	u = u_0 \exp\left[ - \frac{\left(T-T_1\right)^2}{w_t^2}\right]
	\cos\left[\frac{ 2\pi}\lambda\left(T-T_1\right)\right] \label{liper}
\end{equation}
%u_0=-Amp/2
where 
\begin{equation}
T_1=\frac{-\lambda} 2\exp\left( -\frac{Y^2}{w_y^2}\right)
\end{equation}
%w_y=wy(numer)/sqrt(2)
accounts for a transverse initial perturbation,
we have checked numerically (see Fig. 2) that (i) The FCP input transforms into a half-cycle pulse, line soliton solution of KP II equation, whose temporal profile is the sech-square shaped KdV soliton,
plus the accompanying dispersing-diffracting waves, and that
(ii) The initial perturbation vanishes and the line soliton becomes straight again during propagation.

According to the stability analysis \cite{ablo2}, the stabilization of the perturbed line-soliton
occurs if the perturbation is not too large. 
Hence  the essential requirement to evidentiate the above features is the one-dimensional problem of the 
formation of a KdV soliton from a Gaussian input. In principle, this problem is solved  by the 
IST: a wide pulse with high enough intensity is expected to evolve into a finite number of solitons, plus some dispersive waves, called `radiation' in the frame of the IST method.
 This has been addressed numerically in \cite{kdvopt}: it has been seen that, for a FCP-type input, the
amount of radiation was quite important, and the exact number and size  of solitons might depend on the carrier-envelope phase.

 Numerically, the finite size of the computation box and periodic boundary conditions prevent the dispersive wave from being totally spread out. Hence, in order to evidence  numerically the phenomenon mentioned above, we need to reduce the energy of the dispersive part to the lowest level possible, and produce a single soliton.
Very roughly, one can say that each positive half-cycle oscillation with high enough amplitude 
results in one soliton. The initial carrier envelope phase should be close to zero, the main oscillation 
have a large enough amplitude, and the pulse be short enough so that only one peak is emitted.
On the other  hand, the length of the KdV soliton (\ref{kdvsoli}) is  $2/p$, while a half-cycle of the pulse (\ref{liper})  obviously has duration $\lambda/2$. On the other hand, the amplitude of  the KdV soliton (\ref{kdvsoli}) is $6p^2A/B$, to be compared to the amplitude $u_0$ of pulse (\ref{liper}). In a first approximation, the half-cycle matches the KdV soliton if both the durations and  amplitude are the same, hence if
\begin{equation}
u_0=\frac{96A}{\lambda^2B}. \label{opti}
\end{equation}
Numerical computation confirms that relation (\ref{opti}) between the wavelength and amplitude of the Gaussian pulse is roughly optimal. 

\begin{figure}
\begin{center}
\subfigure[ ]{\includegraphics[width=6cm]{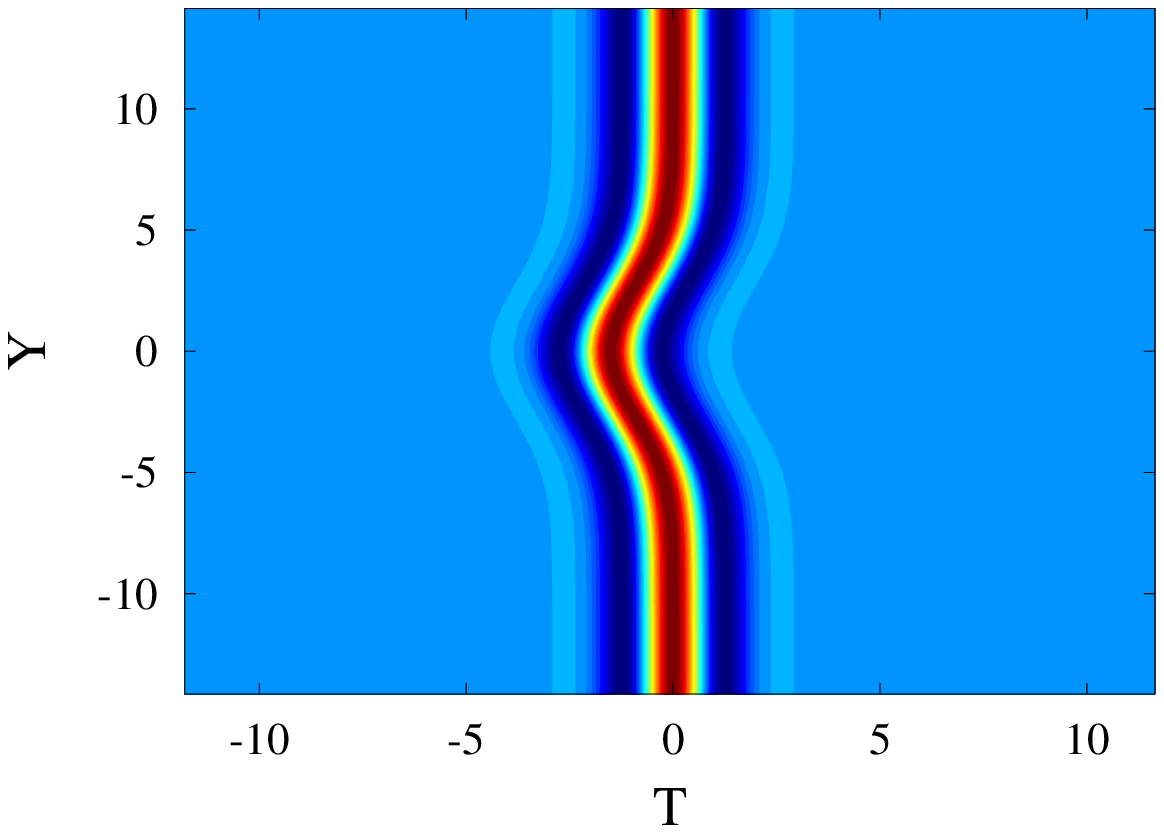}}
\subfigure[ ]{\includegraphics[width=6cm]{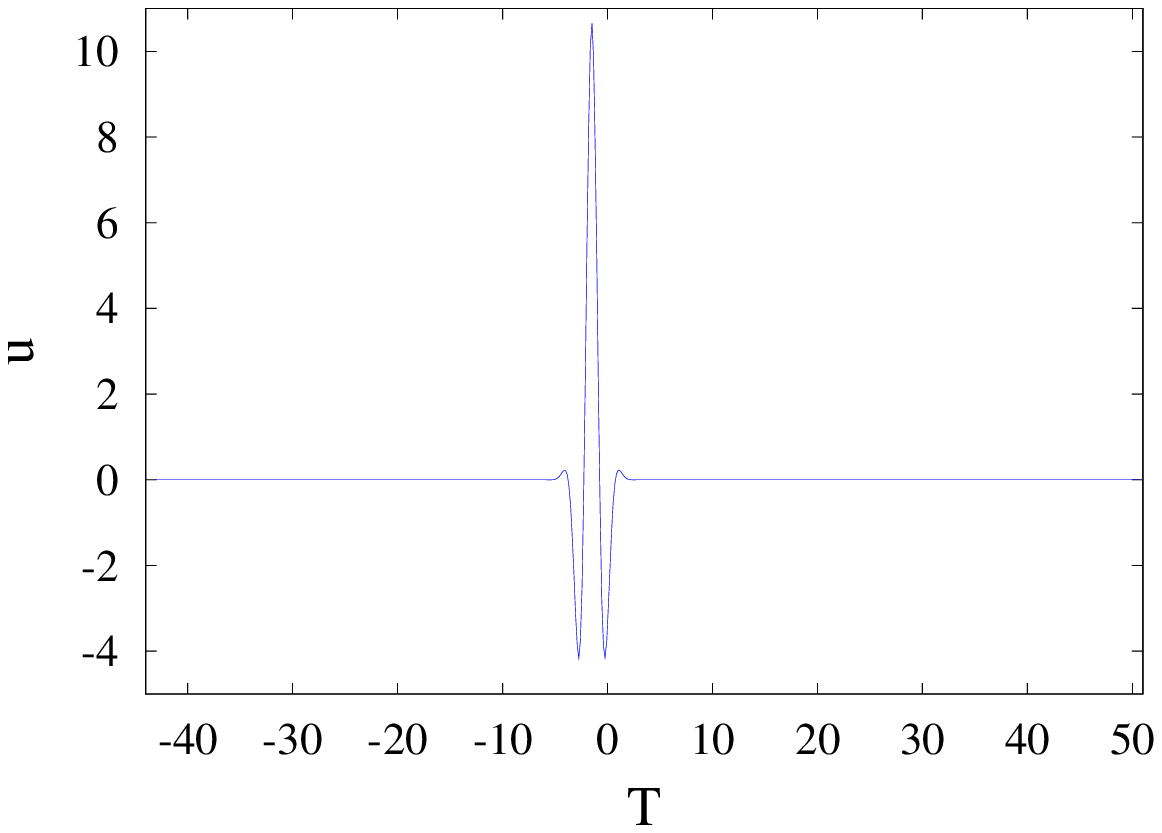}}
\subfigure[ ]{\includegraphics[width=6cm]{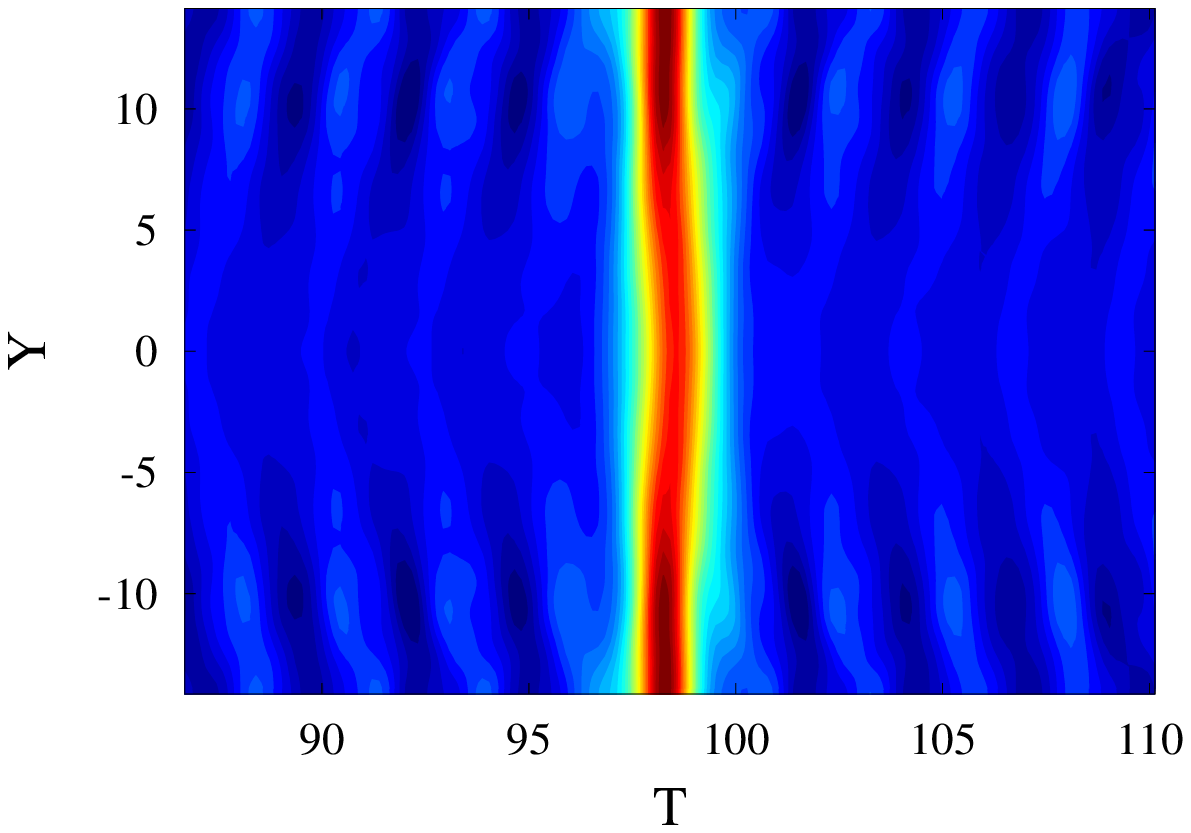}}
\subfigure[ ]{\includegraphics[width=6cm]{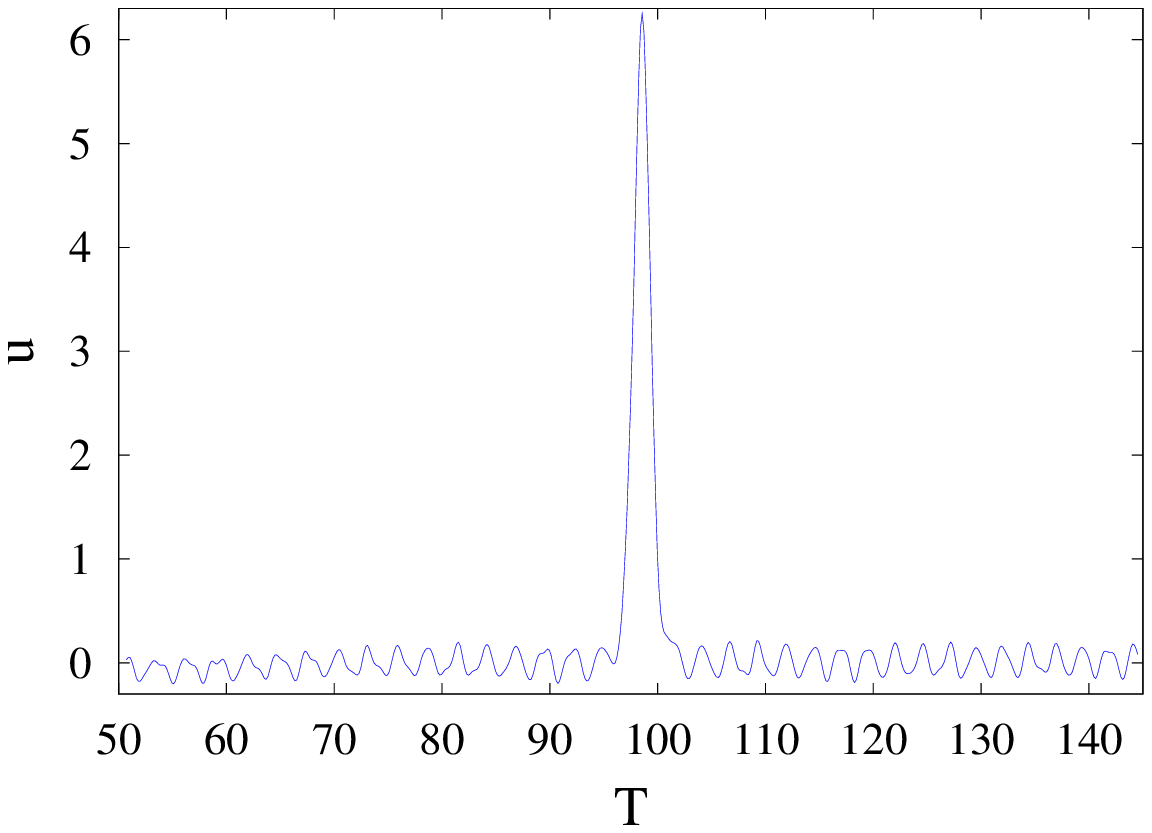}}
\caption{(Color online)
Shape and profile of the line soliton. a-b: input data, c-d: after propagation $Z=15$ (exactly 14.95).
%attention, a corriger u->-U/2,y->y/sqrt(2) FAIT
Parameters are $\lambda=3$, $u_0=10.7$, %-Amp/2=+192/2lam^2
$w_t=1.4$, and $w_y=3.87$.} %2.74*sqrt(2)}
\label{line}
\end{center}
\end{figure}

Recall that our numerical computations are performed using the normalized
 form (\ref{kpr}) of the KP  equation, which corresponds to $A=B=1$. We
observed that, for the particular values of  $\lambda$ and $w_t$ pertaining to Fig. \ref{line},
for $u_0\gtrsim 5$, the soliton emerges from the dispersive wave, while, at optimum (\ref{opti}) ($u_0\simeq 11$), 
the soliton amplitude is about three times the amplitude of the dispersive wave.

A stable plane-wave half-cycle quadratic optical soliton can thus be formed from a FCP input.
Further, the quadratic nonlinearity may restore the spatial coherence of the soliton.

\subsection{KP I}

\begin{figure}
\begin{center}
\subfigure[ ]{\includegraphics[width=6cm]{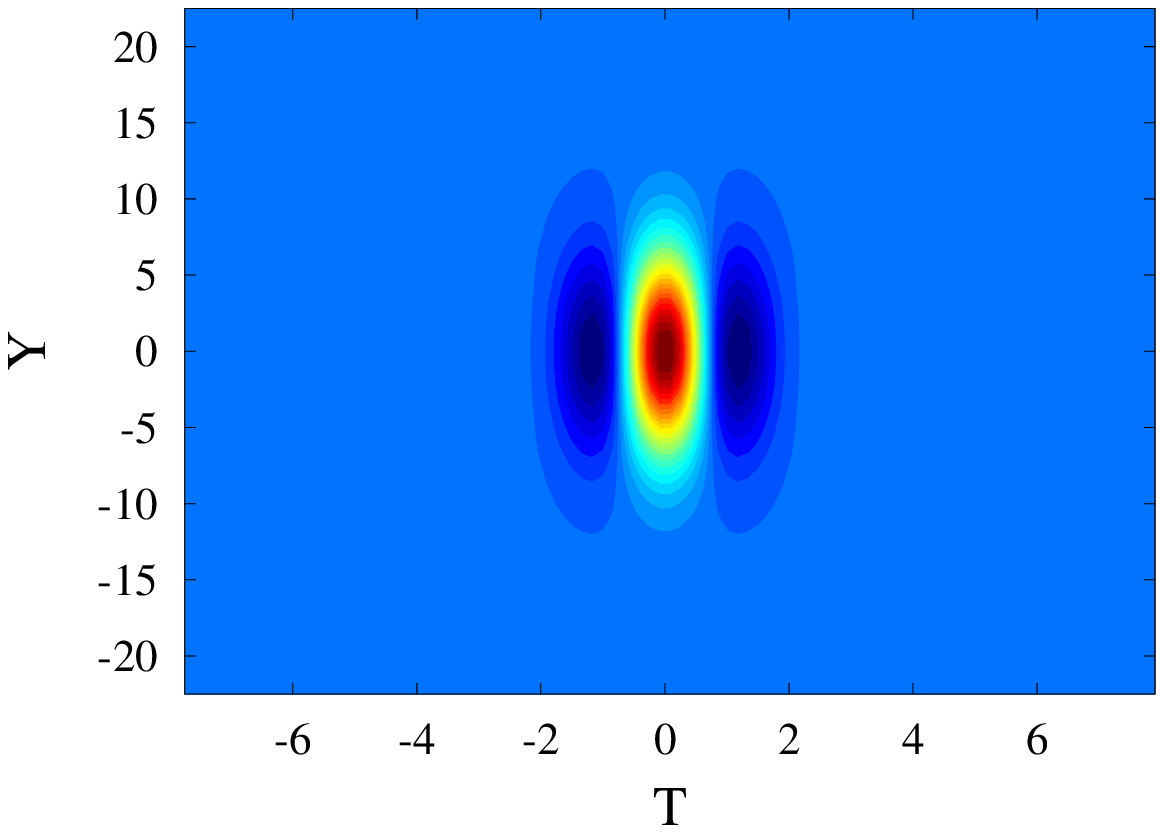}}
\subfigure[ ]{\includegraphics[width=6cm]{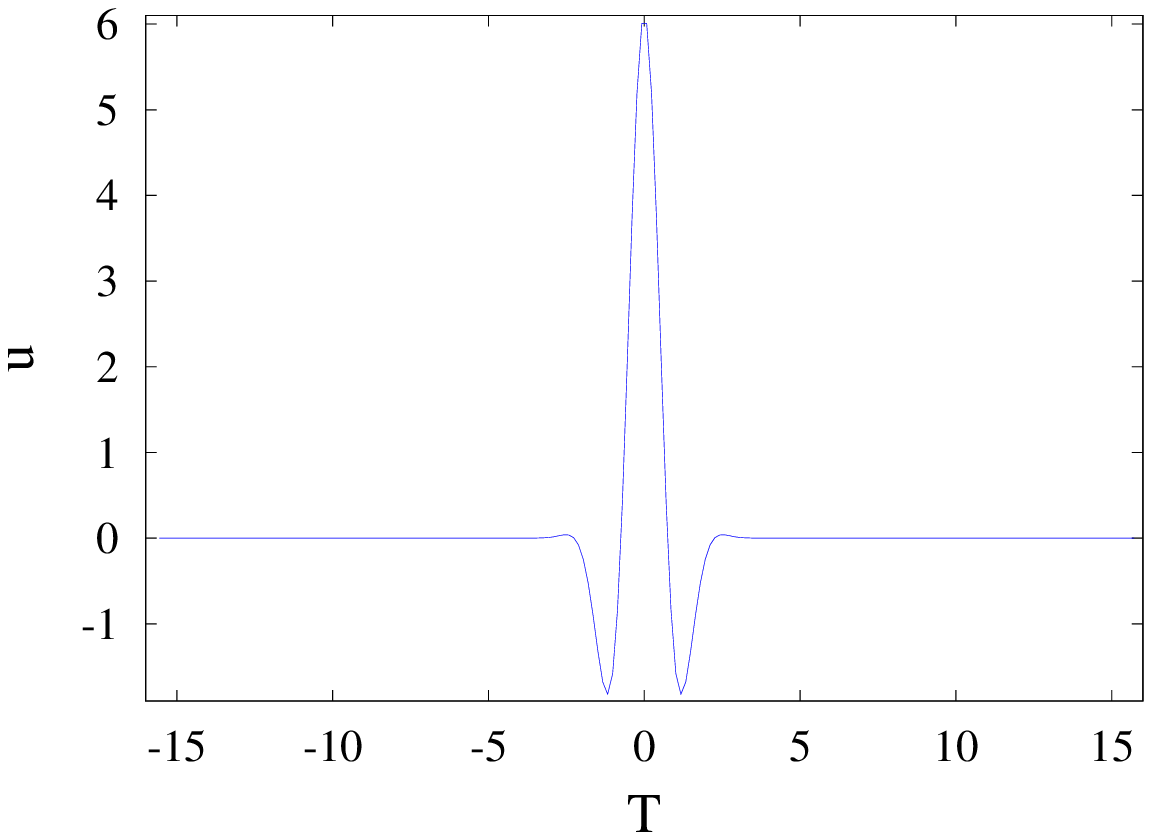}}
\subfigure[ ]{\includegraphics[width=6cm]{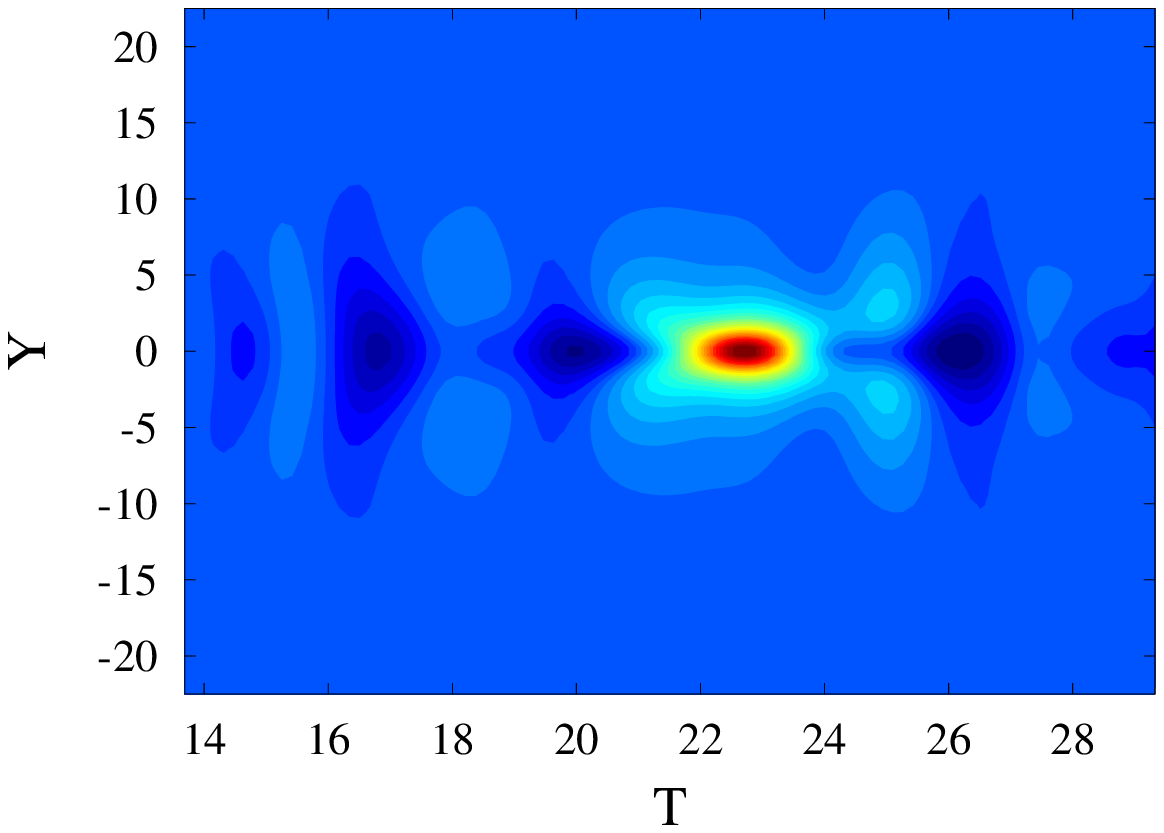}}
\subfigure[ ]{\includegraphics[width=6cm]{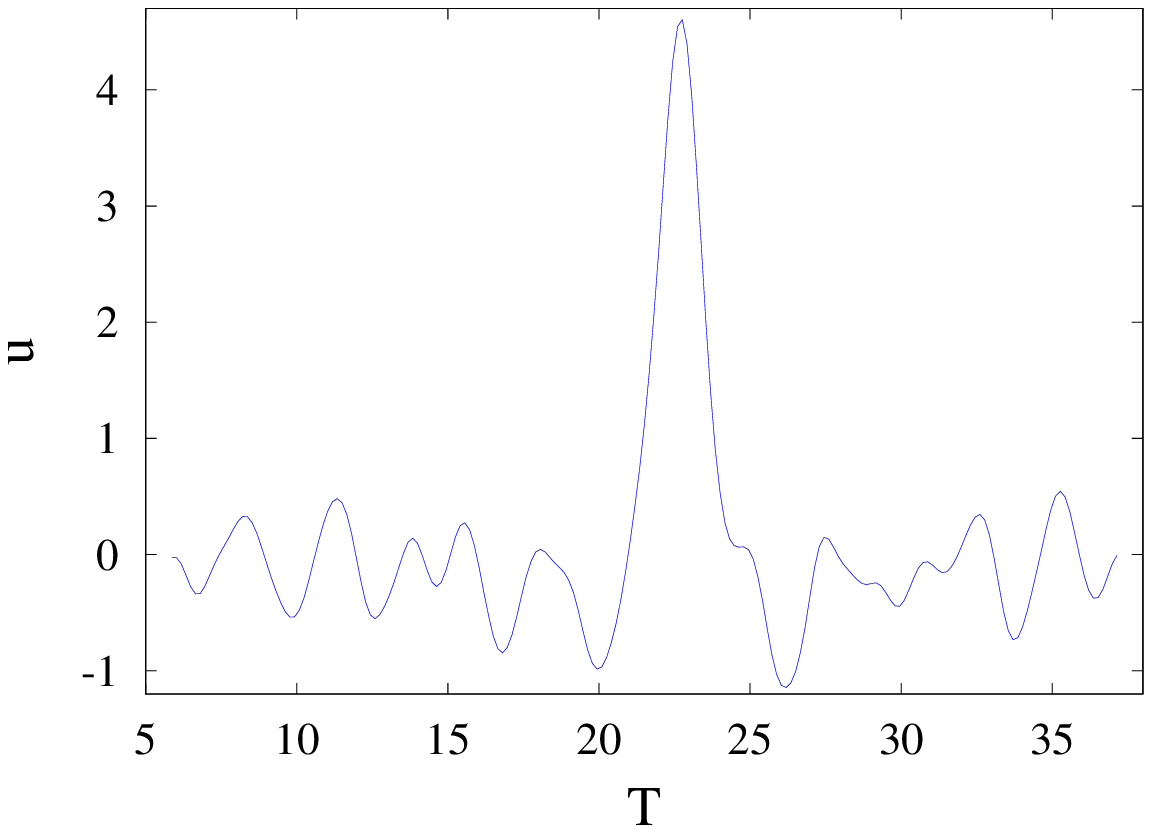}}
\label{figlump}
\caption{(Color online)
Shape and profile of the lump. a-b: input data, c-d: after propagation $Z=15$ (exactly 14.95).
Parameters are $\lambda=3$,
$u_0=6.12$, %Amp/2;Amp =5 *sqrt(6) ;
$w_t= 1.2$, and $w_y=6.36$. %4.5*sqrt(2)
}
\end{center}
\end{figure}

\begin{figure}
\begin{center}
\subfigure[ ]{\includegraphics[width=6cm]{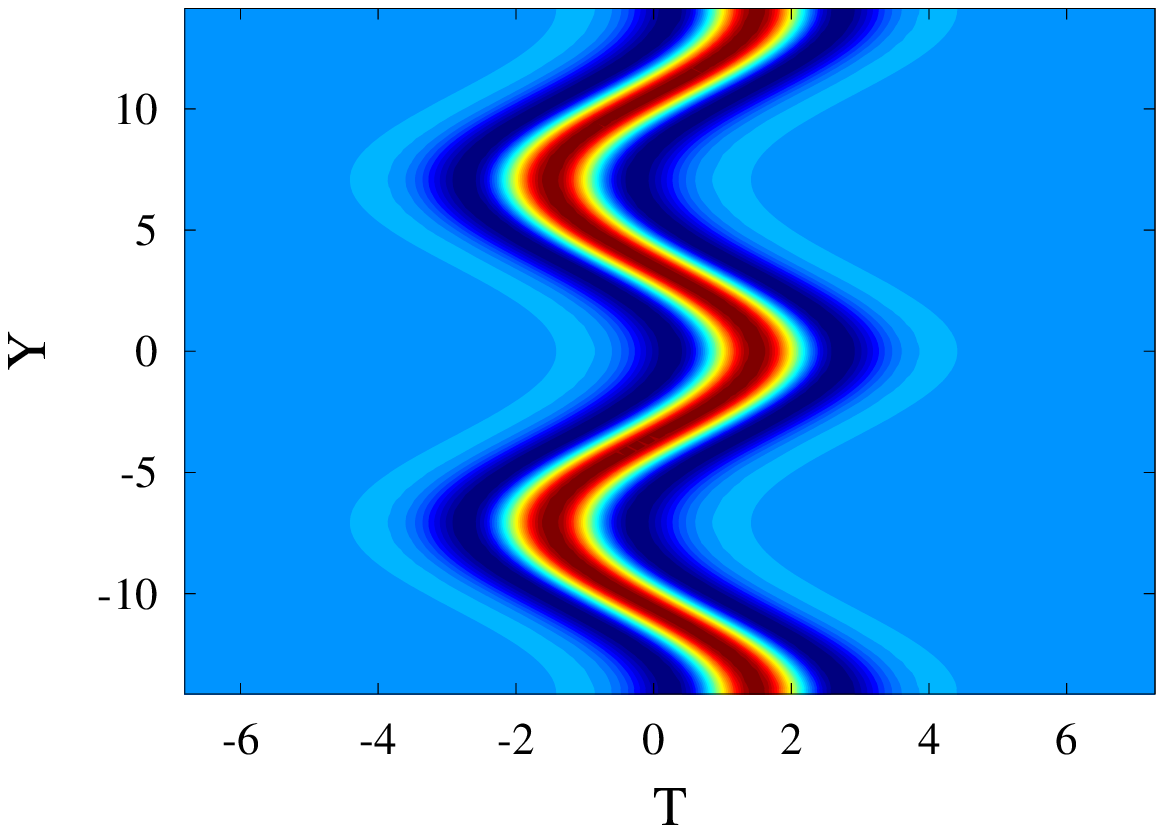}}
\subfigure[ ]{\includegraphics[width=6cm]{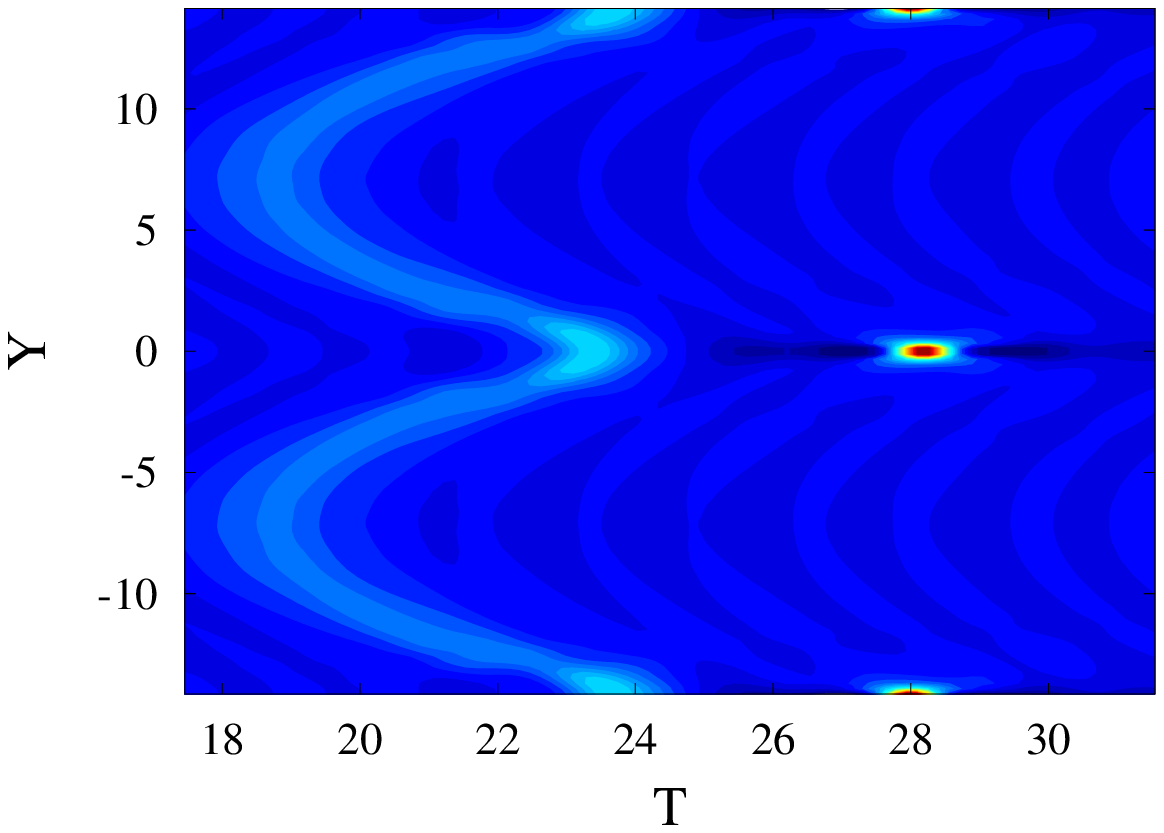}}
\caption{(Color online) Generation of lumps from a perturbed  unstable line soliton.
a : input data, b: after propagation $Z=1.45$.
Parameters are $\lambda=3$,
$u_0=10.7$, %Amp/2;Amp=192/lam^2;
$w_t= 1.4$, and $L=14.14$, 
%per_y=1/2*fenetre_y;%wy=2.74
}\label{snake}
\end{center}
\end{figure}

Let us now assume a quadratic nonlinear medium with anomalous dispersion 
($n''<0$), in which the KP equation (\ref{kdvn1}) is valid.
It reduce to  Eq. (\ref{kpr}) with $\sigma=-1$, i.e.,  
KP I equation. 

We consider an initial data of the form
\begin{equation}
u = u_0 \exp\left( - \frac{T^2}{w_t^2}-\frac{ Y^2}{w_y^2 }\right)\cos\left(\frac{2\pi}\lambda T\right),
\end{equation}
with arbitrary values of the initial amplitude $u_0$, width $w_y$, and duration $w_t$ of the pulse.
The evolution of the input FCP into a lump is shown on Fig. 3; notice the remaining dispersive waves.
The fact that they  seem to be confined in the $Y$-direction is a numerical artifact, due to the fact that absorbing boundary conditions have been added in the $Y$-direction only. Hence the diffracting part of the dispersive waves rapidly decreases as if the width of the numerical box were infinite, while the dispersive waves which propagate along the $Z$-axis remain confined in the numerical box (with length 160), and hence decays slower as it would happen for a single pulse.

Figure 4 shows the decay of a perturbed unstable line soliton of KP I equation into lumps. 
The input is given by Eq. (\ref{liper}), with the transverse perturbation
\begin{equation}
T_1=\frac{-\lambda}2\cos\left(\frac{2\pi}L Y\right),
\end{equation}
where the wavelength $L$ of the transverse perturbation is chosen so that $2L=28.3$ is 
the width of the numerical box, in accordance
 with the periodic boundary conditions used in our numerical computation. %fenetre_y*sqrt(2)
This transversely perturbed line-soliton, in contrast with the case of KP II for normal dispersion shown in Fig. 2,
does not recover its initial straight line shape and transverse coherence, but breaks up into localized lumps. 
Notice that the conditions on the parameters needed to obtain a result comparable to that shown in Fig. 4 
are much less restrictive than the conditions required for the numerical observation of 
line solitons; see Fig. 2 and the detailed discussion given above on the optimal conditions of generation of line solitons of KP II equation.
Hence spatiotemporal quadratic solitons may form from a FCP input,
their transverse focusing follows from any transverse perturbation of the incident plane wave.

\section{Conclusion}

In conclusion, we have introduced a model beyond the slowly varying envelope approximation of the nonlinear Schr{\"o}dinger-type evolution equations, for describing the propagation of (2+1)-dimensional spatiotemporal ultrashort optical solitons in quadratic nonlinear media. Our approach is based on the Maxwell-Bloch equations for an ensemble of two level atoms. We have used the multiscale approach up to the sixth-order in a certain small perturbation parameter and as a result of this powerful reductive perturbation method, 
two generic partial differential evolutions equations were put forward, namely, (i) the Kadomtsev-Petviashvili I equation (for anomalous dispersion),  and (ii) the  Kadomtsev-Petviashvili II equation (for normal dispersion).
Direct numerical simulations of these governing partial differential equations show the generation of both stable line solitons (for the Kadomtsev-Petviashvili II equation) and of stable lumps (for the Kadomtsev-Petviashvili I equation). Moreover, a typical example of the decay of the perturbed unstable line solitons of the Kadomtsev-Petviashvili I equation into stable lumps is also given.

The present sudy is restricted to (2+1) dimensions, however, due to the known properties of the Kadomtsev-Petviashvili equation in (3+1) dimensions \cite{inf95a,sen98a}, analogous behavior is expected in three dimensions, i.e., stability of the spatial coherence of a plane wave 
few-cycle quadratic soliton for normal dispersion, and spontaneous formation of spatio-temporal few-cycle `light bullets' in the case of anomalous dispersion.

\end{document}